\documentclass[runningheads]{llncs}
\usepackage{graphicx}
\usepackage{url}

\usepackage{booktabs}
\usepackage [english]{babel}
\usepackage [autostyle, english = american]{csquotes}
\MakeOuterQuote{"}

\usepackage[numbers,sort&compress]{natbib}

\begin{document}
\title{Structuralist analysis for neural network system diagrams}
%
%
\author{Guy Clarke Marshall\inst{1}, Caroline Jay\inst{1}, and Andr\'e Freitas\inst{1,2}}
\authorrunning{GC Marshall et al.}
\institute{Department of Computer Science, University of Manchester, Manchester, UK \\ \email{guy.marshall@postgrad.manchester.ac.uk, \\ \{caroline.jay, andre.freitas\}@manchester.ac.uk}
\and 
Idiap Research Institute\\
Rue Marconi 19, Martigny, 1920, Switzerland\\
}
\maketitle              
\begin{abstract}
This short paper examines diagrams describing neural network systems in academic conference proceedings. Many aspects of scholarly communication are controlled, particularly with relation to text and formatting, but often diagrams are not centrally curated beyond a peer review. Using a corpus-based approach, we argue that the heterogeneous diagrammatic notations used for neural network systems has implications for signification in this domain. We divide this into (i) what content is being represented and (ii) how relations are encoded. Using a novel structuralist framework, we use a corpus analysis to quantitatively cluster diagrams according to the author's representational choices. This quantitative diagram classification in a heterogeneous domain may provide a foundation for further analysis.

\end{abstract}
%
%
%




\section{Introduction}




Currently, there is no consistent model for visually or formally representing the architecture of neural networks (NNs). This lack of representation brings interpretability, correctness and completeness challenges in the description of existing models and systems engineering. In the context of scientific communication, most approaches and systems today are described by a combination of arbitrary diagrammatic elements, algorithms, formulae and natural language descriptions. In this paradigm, there is little consistency on abstraction levels or notation. From the perspective of scientific practice, these limitations challenge dialogue, transparency and reproducibility.

We will show that authors of applied NN systems papers could be getting more communicative value from their diagrams. Focusing on the diagrammatic representations used, it seems the author's notion of relevance is skewed, resulting in missing context and content. Diagrams often feature partial information on, for example, the key novelty or contribution the paper is articulating. This is unlike other disciplines such as mechanical, electronic, or structural engineering, where standardised and complete diagrams are utilised.


This short paper is an empirical analysis based on semiotic theory, introducing a semiotic viewpoint to the analysis of NN diagrams used at leading Natural Language Processing (NLP) conferences, applying a novel structuralist approach, inspired by structural linguistics \citep{jakobson2010fundamentals} to survey a corpus of NN diagrammatic representations. Our contributions are:
\begin{itemize}
    \item To introduce diagrammatic structuralism as a method for analysing diagrammatic semiotics.
    \item To show the heterogeneity of neural network diagrams through an empirical corpus-based analysis.
    \item To identify clusters of diagrams through quantitative analysis.
\end{itemize}
%

\section{Related work}

\subsection{Scientific diagrams}

In an interview study involving academics about NN system diagrams, \citet{marshall2020researchers} found differing views on whether precision was meaningful: "If you have four, it means you have four" contrasts with "it cannot be four dimensions, it must be much more than that."

Specific semiotic principles for diagrams can be found in other technical applications. In the domain of Business Process Mapping, it was found that the shape of the graphic symbols improved ease of understanding more effectively than colour or number of symbols \citep{Gabryelczyk2017DoesNotation}. In a small study of mechanical aviation diagrams, \citet{Kim2010EvaluatingTasks} found that highlighting/shading to identify key components was more usable and useful, compared to other visual techniques such as "zooming in". Additionally, \citet{Heiser2006ArrowsDiagrams} showed in mechanical diagrams that arrows added to static structural (parts and relations) diagrams increase the capability of the representation to convey function (temporal, dynamic, or causal process). 


\subsection{Diagrammatic representations}
According to Peirce \cite{Peirce1998CharlesWritings}, the concept of a diagram as an icon is malleable to Kant's concept of Schema (the procedural rule by which a Kantian category is associated with a sense impression). Applying Saussurian semiotic thinking beyond the linguistic domain, in a diagram we might wish to minimise the distance between the signifier and the signified. There is agreement that, at the holistic and iconic level, a diagram should attempt to represent the thing it is representing. \citet{Tylen2014DiagrammaticInsight} concisely summarise that diagrams:
\begin{itemize}
    \item are external representational support to cognitive processes \citep{Clark1998TheMind}.
    \item make abstract properties and relations accessible \citep{Hutchins1995HowSpeeds}.
    \item can be in a public space, therefore enabling collective and temporally distributed forms of thinking \citep{Peirce1998CharlesWritings}.
    \item are manipulated in order to profile known information in an optimal fashion.
\end{itemize}

Diagrams and text, even if together in the same media, are different modalities. Due to the prevalence of their combined usage, when referencing diagrams we assume dual coding of text and graphics as part of our working definition of a diagram. We build on \citet{Morris1938FoundationsSigns} by partitioning semiotics, the field examining the process by which something functions as a sign, into the following categories:
\begin{itemize}
    \item \textit{Semantics}: relation between signs and the things to which they refer, or their meaning
    \item \textit{Syntactics}: relations among or between signs
    \item \textit{Pragmatics}: relation between signs and sign-users \citep{Morris1938FoundationsSigns}
    \item \textit{Empirics}: the statistical analysis of repeated use of signs, adapted from the Information Systems domain \citep{Stamper1987Semantics}
\end{itemize}

\section{Neural network diagram empirics}
\label{section:NN}
\subsection{Context}

Neural approaches are popular in AI research, and many of the diagrammatic design decisions are centred on the machine-learning mechanisms. We performed a corpus-based analysis on the use of diagrams to represent NNs, with NLP as an application domain. 


This study applies an empirical approach to diagrammatic representations, inspired by structuralist linguistics, in order to uncover the potential implicit semiotic processes and laws. This approach has a potential shortcoming in revealing only partial truths \citep{Rutherford2016HowMatters}, but allows us to access a complex conceptual space. By analogy to Saussure's \citeyear{Saussure2011CourseLinguistics} linguistic concept of \textit{parole} (speech as concrete instances of the uses of language), the proposed \textit{diagrammatic structuralism} purely considers the sign itself and does not include the act of using it. To assess the similarity of diagrams from a visual perception standpoint, we should consider the overall structure rather than substructures \citep{Goldmeier1972SimilarityForms.}. Following Saussure, the American Structuralist school (e.g. \citet{Bloomfield1984Language}) advocate a mechanistic approach to the analysis of linguistics, which we apply to NN diagrams. We believe this approach to be pragmatic and not excessively reductionist, as the establishment of common graphical elements is beneficial for understanding what is being articulated.



\subsection{Method}

\begin{table}[tbp]
    \caption{Surveyed Papers}
    \label{tab:NN_tableSurveyed}
    \vskip 0.15in
    \begin{center}
    \begin{small}
    \begin{sc}
    \centering
    \begin{tabular}{lrrr}
        \toprule
        Source & Papers & Sampled & Diagrams\\
        \midrule  
        COLING 2016 & 338 & 16 & 6\\
        COLING 2018 & 330 & 16 & 8\\
        EMNLP 2016 & 275 & 9 & 4\\
        EMNLP 2017 & 323 & 14 & 6\\
        NAACL 2016 & 181 & 8 & 4\\
        NAACL 2018 & 332 & 14 & 7\\
        NIPS 2016 & 569 & 14 & 2\\
        NIPS 2017 & 679 & 23 & 3\\
        \bottomrule
    \end{tabular}
    \end{sc}
\end{small}
\end{center}
\vskip -0.1in
\end{table}
In order to map the current state of diagrammatic representations of NNs, we conducted systematic literature analysis, categorising NN diagrams based on Diagrammatic and Content features (see \citet{marshall2021resources} for a complete list, including observed frequencies). This partition is inspired by Jakobson's \cite{jakobson2010fundamentals} "axis of selection" and "axis of combination" applied in structural linguistics. Prior to the sampling, we identified repeated concepts, divided into "Content" and "Diagrammatic" primitives, a diagrammatic equivalent of Jakobson's "selection" and "combination". The categorisation is in this way because the sets of symbols and their relationships together can also form a symbol themselves, in the same way that the set of low-level symbols come together to for the diagram symbol. This intertwining means it is not possible to distill semantic and syntactic information independently, as argued by Jakobson for linguistics.

We then sampled and applied our coding based on content and diagrammatic features (see Table \ref{tab:NN_tableSurveyed} for content, diagrammatic features table omitted for brevity), analysing the results.

With a corpus-based approach, we randomly sampled papers and manually extracted 40 diagrams (see Table \ref{tab:NN_tableSurveyed}). Where multiple diagrams were identified, we have taken the diagram including the most features. We excluded Computer Vision venues to avoid complication with the visual nature of their input and high specificity of their architectures, and as such this is not representative of all scholarly NN diagrams.

The diagrams were assessed by one NN practitioner according to pre-defined categories covering Content (what features have been selected for diagrammatic representation) and Diagrammatic concepts (their visual and relational encoding).


\subsection{Analysis}



The low occurrence of data-centric content features such as operations indicates that these diagrams are not describing transformations the data undergoes, but instead prioritise the NN system, particularly the layers, dimensionality, input and output.


In order to extract 40 diagrams, we sampled 114 papers. Table \ref{tab:NN_tableSurveyed} details the paper counts from each conference. From papers at these recent top neural network-centric venues, we expect that at least 25\% will contain NN diagrams (with 99\% confidence, due to our sampling of 40 diagrams from 114 sampled papers from a total population of 3027). Note that the sampling was not filtered for NN content, and the prevalence of NN diagrams underscores the community requirement for effective NN diagrams.

\begin{figure}[htbp]
    \centering
    \includegraphics[width=0.8\textwidth]{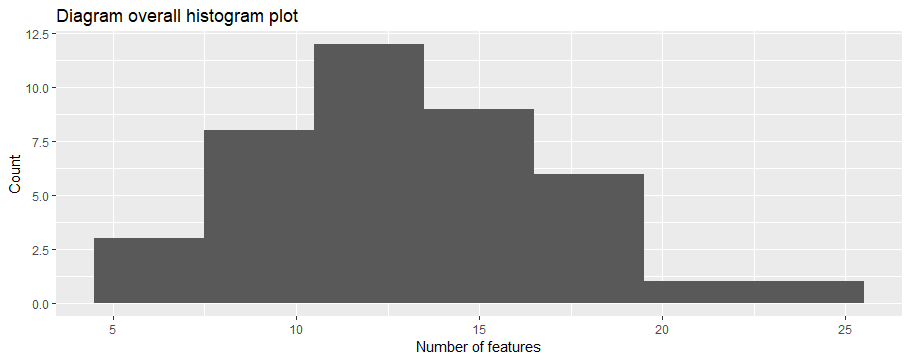}
    \caption{Distribution of the total number of features in each diagram}
    \label{fig:OverallFeatures}
\end{figure}

Most diagrams are representing the same content. In 82.5\% of papers, the role of NN diagrams in the sample was to communicate the function of the NN, rather than context, the whole system, or a component. In the sample, \textit{no diagram included identical Diagrammatic features}. Two of the sampled diagrams shared the same Content features (lower semantic content, NN-only diagrams, with an output and an explicit example). Fig. \ref{fig:OverallFeatures} shows the quantity of semiotic elements according to our schema, which were found to be normally distributed (13.07) and have a large standard deviation (3.96). Details are available \citep{marshall2021resources}. Some highlights: 
\begin{itemize}

    \item 92.5\% used directional arrows, and 42.5\% dotted arrows. Given the utility of arrows to provide a sense of functional composition, this uniformity is as expected. 
    \item 60\% had a caption over one sentence long, perhaps suggesting the authors struggled to get the relevant content into the diagram.
    \item 60\% used an ellipsis in order to indicate "incomplete sets of objects." 
    \item 47.5\% included labels on neural layers. 
    \item 42.5\% made use of an explicit example input. 
    \item 52.5\% did not indicate dimensionality.
    \item 27.5\% indicated the key features by boundary lines, shading or labeling. 
    \item 10\% indicated data resources of any kind, and a different 10\% utilised symbolic mathematical tensor operation. 
    \item None used UML, SysML, OML or other existing formal diagrammatic language.
\end{itemize}

\begin{figure}[htbp]
    \centering
    \includegraphics[width=0.8\textwidth]{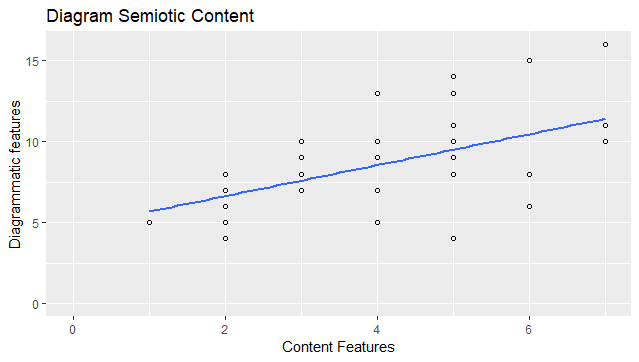}
    \caption{Surveyed Semiotic Content of Diagrams}
    \label{fig:NNSurveySemiotics}
\end{figure}

Fig. \ref{fig:NNSurveySemiotics} shows the relationship between NN Content and Diagrammatic features. Diagrams with additional feature content are more likely to have additional diagrammatic features. It may be that there is variable author diagramming efficiency, or that these systems are intrinsically more complex in both dimensions. 

The study found little evidence of standardisation. In the language of Peirce, the level of heterogeneity in representing such similar \emph{objects} implies that it is more appropriate to consider the components of NN diagrams as \emph{symbols} rather than \emph{icons}, since conventions appear to be utilised over physical resemblance of system or data structures. Despite this heterogeneity, common themes have been identified in order to classify and group diagram types. Based on \textit{k-means clustering} on the 69-dimensional space created by applying diagrammatic structuralism (24 content and 45 diagrammatic), together with manual adjustment, we identified four potential NN diagram classes:

\begin{enumerate}
    \item \textit{Visual} using coloured circles to represent tensors, mostly left-to-right shape, mostly including an input and output example, a short caption, and featuring an embedding.
    \item \textit{Mathematical} with more mathematical notation, often with an upward linear shape.
    \item \textit{Lightweight} with less content and fewer visual encoding mechanisms, relying on text, often with labelled layers, and sometimes in a Block diagram format.
    \item \textit{Unorthodox} not fitting into the above groups.
\end{enumerate}

Applying this classification manually, we identified 8 Visual, 4 Mathematical, 12 Lightweight and 16 Unorthodox diagrams. This classification may help structure future research.

\subsection{Limitations}
\begin{itemize}
    \item The dimensions were chosen to have a good coverage of important features in the space. Whilst this was intended to be comprehensive, because the dimensions were identified before the sampling (a methodological choice), several features were inadvertently omitted. This was partly due to the heterogeneity of the representations and included colour relationship, the presence of keys, the usage of standard notation in non-standard ways, presence of ambiguous syntactics.
    \item It would be an improvement to have multiple annotators. However, due to the reasonably unambiguous nature of the assessment, this was not deemed critical in this early stage of analysis. 
    \item We do not include bibliometrics such as citation count due to the small sample size and the sensitivity in adjustment for different publication venues. 
\end{itemize}


\subsection{Discussion}
The consequence of variable signification methods within the same domain is that readers are required to switch modalities even between very similar papers. Readers may also be unable to access the information they require, as the diagrammatic signifier may not contain some important Content-related aspects of the signified system. 

We found that additional visual objects were correlated with additional diagrammatic grammar features. This suggests that either the expressions were more complex and required both types of encoding, or that some author were more likely to use a larger variety of encoding objects generally (not discriminating between objects and relations). We have not shown that this heterogeneity is a problem per se, but it does differ from many other Computer Science domains where diagrammatic representation of systems is more standardised.

\citet{grice1989studies} distinguishes between "what is said" and "what is implicated". In this domain, the figure caption text usually refers to a system rather than a scholarly contribution. It may be that in these diagrams we are observing an ambiguity between "what is drawn" and "what is implicated", even for the author's own mind. The system diagram may be functioning as a Peician index for a scholarly contribution, rather than a system, the functioning of that system, or the manipulation of data. This ambiguity introduces complexity in identifying the object of the diagram. Despite this ambiguity, a semiotic lens has allowed quantitative exploration of diagram components and poses further questions about the intent of the authors.

\section{Conclusion}

Using a structuralist framework we gathered, sampled and analysed neural network diagrams from a variety of recent neural network-centric conferences, using this to demonstrate aspects of the heterogeneity of diagrammatic representations employed. Using a k-means clustering algorithm on the high-dimensional visual semiotic space, we derived clusters of NN diagram types.

Beyond diagram classification and hetergeneity quantitification, this work aims to encourage increase attention on scholarly diagramming practices, which may help authors to better fulfil their communicative intent.

\renewcommand{\bibsection}{\section*{References}}
\bibliographystyle{splncs04nat}
\bibliography{bib}
\end{document}